# Sequence adaptive field-imperfection estimation (SAFE): retrospective estimation and correction of $B_1^+$ and $B_0$ inhomogeneities for enhanced MRF quantification

Mengze Gao[1], Xiaozhi Cao[1,2], Daniel Abraham[2], Zihan Zhou[1], Kawin Setsompop[1,2]

[1] Department of Radiology, Stanford University, Stanford, CA, USA

[2] Department of Electrical Engineering, Stanford University, Stanford, CA, USA

INTRODUCTION:

Magnetic Resonance Fingerprinting[1] (MRF) is a widely-used efficient multiparameter mapping approach. However, significant reduction in accuracy and repeatability of the estimated tissue parameters can occur as a subject-and-scan-specific $B_1^+$ and $B_0$ field inhomogeneity[2,3]. $B_1^+$ and $B_0$ calibration scans can be acquired to correct this issue, but cost of added scan time and inability to be applied retrospectively to previously collected data (as these field inhomogeneity vary between scan sessions). $B_0$ estimation/correction via deep learning has been proposed[4,5] for correction of non-cartesian acquisition. However, field map estimation and correction from highly under sampled time-series MRF data has not been attempted. Herein, we proposed a sequence adaptive deep learning framework **(SAFE)** to jointly estimate $B_1^+$, $B_0$, distortion free T1, T2 and Proton Density (PD) maps from subspace coefficients in 3T MRF. Moreover, we demonstrate that our technique is seamlessly adaptable to any 3T MRF sequence and provide accurate field maps with no additional training data.

METHODS:

**Deep-learning-based B1 and B0 estimation:** SAFE estimates field maps directly from the MRF subspace coefficient maps, with auxiliary task of generating ($B_1^+$ and $B_0$ corrected) parameter maps added to improve the performance of the network. The structure is inspired by the U-Net[9], SSIM was employed on the brain-masked area (parenchyma). Due to the smoothness of $B_1^+$ and $B_0$, we down-sampled inputs to 2mm for training.

**Sequence-agnostic capability:** SAFE relies on a set of previously-acquired T1, T2, PD, $B_1^+$, $B_0$ quantitative maps as training data, that can be acquired using any quantitative imaging approach. This data is then used to simulate the reconstructed subspace coefficient maps of the MRF sequence of interest for use as input to the network, where such simulation carefully accounts for $B_1^+$ and $B_0$ effects on the coefficient maps through use of $B_1^+$-corrected dictionary model and spatial blurring of $B_0$ in the spiral time-segmented modelling.

Data Acquisition all *in vivo* data were acquired using GE Premier 3T

system with 32ch reception.

<u>Calibration scan</u> The ground-truth $B_1^+$, $B_0$ maps were collect via Physical[6] sequence with 1mm isotropic resolution.

<u>MRF acquisition</u> Data were acquired on three unique MRF sequences, all utilizing the 3D-SPI-MRF[7] acquisition. The flip-angle-trains for the three sequences (seq1,2,3) are shown in Figure 2A; with seq1 using an acquisition train with 500 TRs and 5.38ms spiral readouts, while seq2 and 3 uses 400TRs and 9ms spiral readouts. 7 healthy adults were scanned with seq to provide high-quality gold-standard 1mm isotropic resolution whole-brain data. Data were collected on two additional subjects using seq2 and 3, along with additional calibration scans for validation.

<u>Ground truth generation for training data synthesis</u> Subspace coefficient reconstruction and dictionary matching were performed to extract ground truth parameter maps from seq1 data. $B_0$ and $B_1^+$ correction is performed via MFI[8] and $B_1^+$-corrected dictionary matching respectively.

<u>Application</u> In addition to validation performed on healthy adult volunteers, we applied our network on on 8–13-year-old children's data acquired at a collaborating site using seq1 from a longitudinal study for brain development .

RESULTS:

Fig3 shows representative field predictions and parameter estimates for seq1 obtained from the network in comparison to ground truth maps. The predicted field maps have high correspondence to ground truth maps, with some smoothing effect on $B_1^+$ map compared to ground truth obtain via the PhysiCal sequence.

Fig4 shows the importance of $B_1^+$ and $B_0$ corrections, in mitigating $B_0$ blurring artifacts and T2 bias in region of low $B_1^+$. Here, the incorporation of B1 information caused a reduction in the estimated T2 value by 24.2% in the highlighted region.

Fig5 shows the application of SAFE on sequence 2 and 3, where high quality field predictions are also obtained.

DISCUSSION:

SAFE exhibits the capability for high-quality estimation of $B_1^+$ and $B_0$ maps across different MRF sequences. While using the convolutional layer allowed for capitalizing on spatial relations between voxels, larger receptive field improve network accuracy due to field map dependency on proximal regions.

CONCLUSION:

We demonstrate the ability of SAFE for robust and accurate $B_1^+$ and $B_0$ field estimations on arbitrary MRF sequences. It should be feasible to estimate other field/system imperfections and apply to other quantitative imaging sequences such as EPTI[10] and MR multitasking[11]. For applications to other magnetic field strengths and/or other organs, to get robust results, it will be important to train on $B_1^+$, $B_0$, T1, T2 and PD that are in distribution to those applications, which can be obtained similarly here a good quantitative sequence for the specific application. Data augmentation approaches here could also further improve the robustness of our framework.

ACKNOWLEDGEMENT:

This work was supported by NIH research grants: R01MH116173, R01EB019437, U01EB025162, P41EB030006, R01EB033206, U24NS129893.

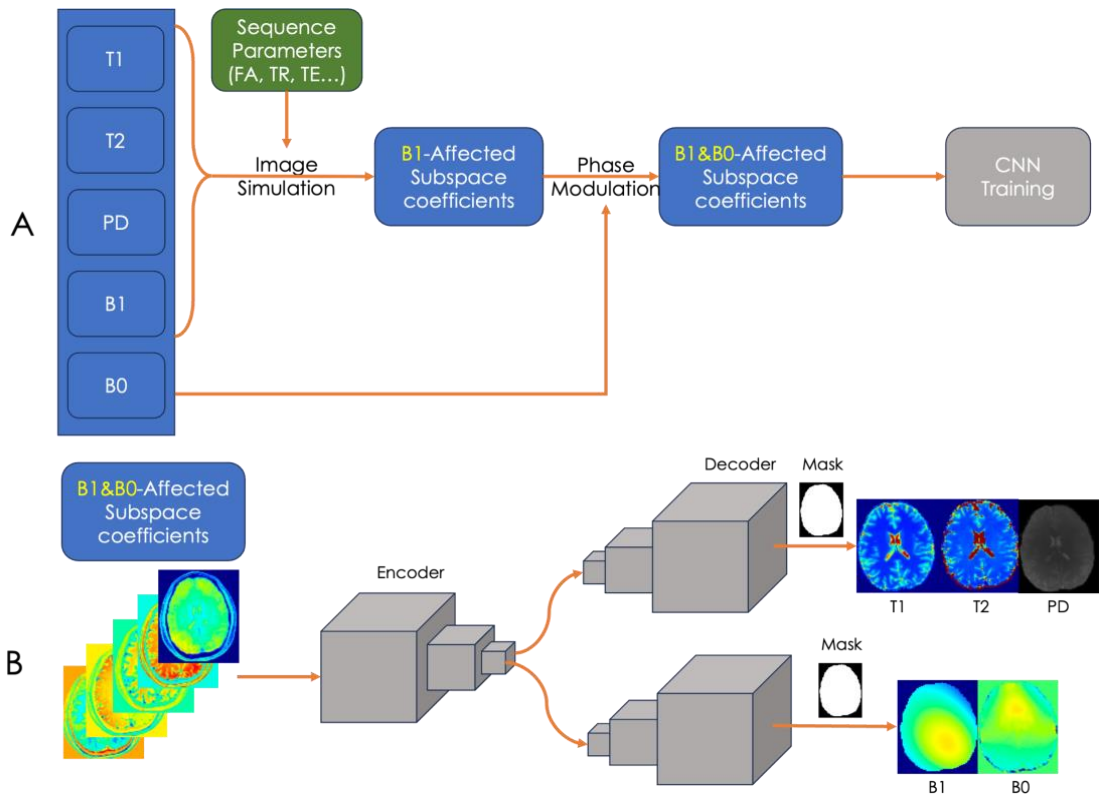

figure 1. (A) Simulate $B_1^+$-corrupted subspace coefficients based on ground truth quantitative parameter maps for the target MRF sequence with specific scan parameters (TR, TE, FA) and trajectories. By adding phase modulation onto the resulting k-space data based on $B_0$ map, the $B_1^+$ & $B_0$-corrupted subspace coefficients were synthesized as the input for network training.

(B) Detailed network training structure. The network block consists of convolutional, instance normalization and activation layers, with skip connections between encoders and decoders.

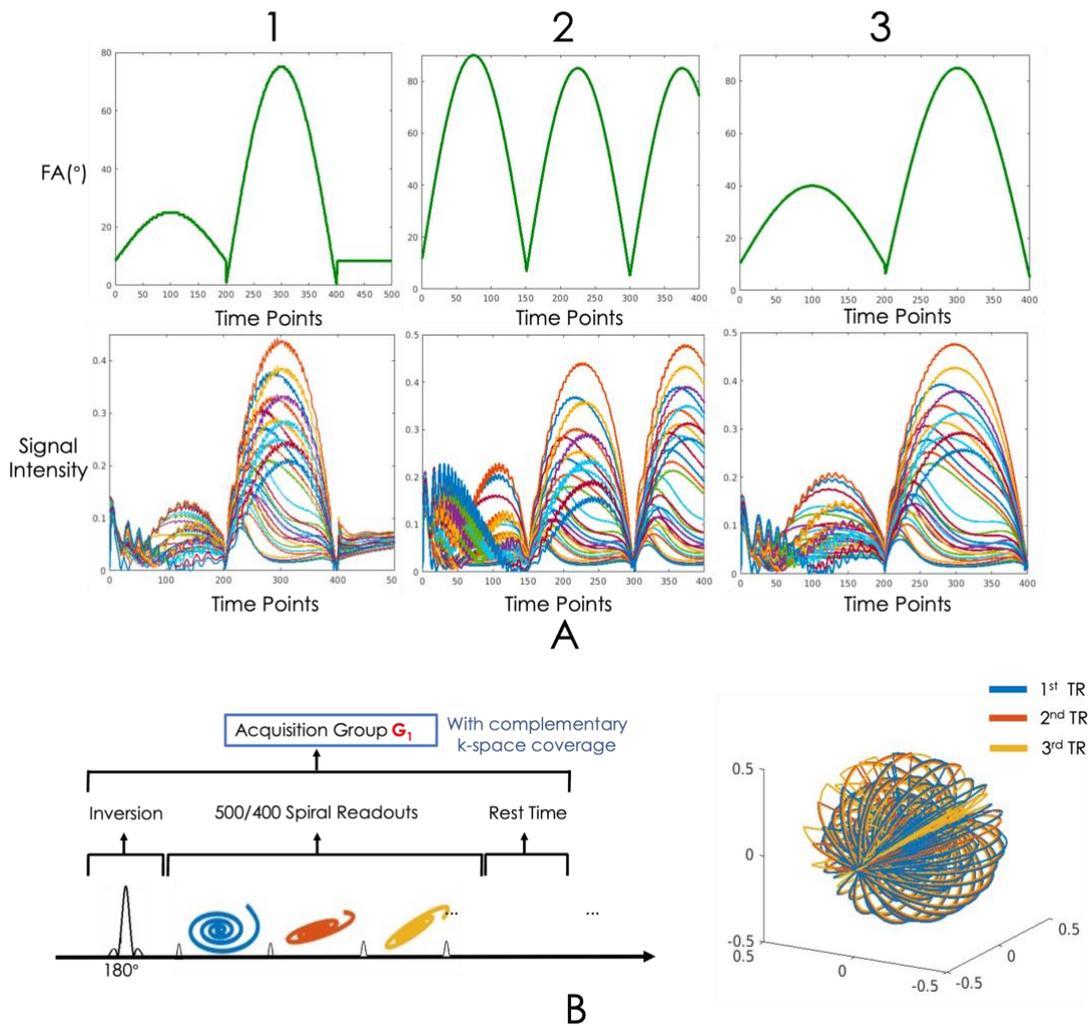

Figure 2. (A) MRF sequence with different scan parameters (TR, TE, FA) and trajectories. (B) Sequence diagram and k-space coverage (for the first 3 TRs) of the first MRF sequence shown in (A).

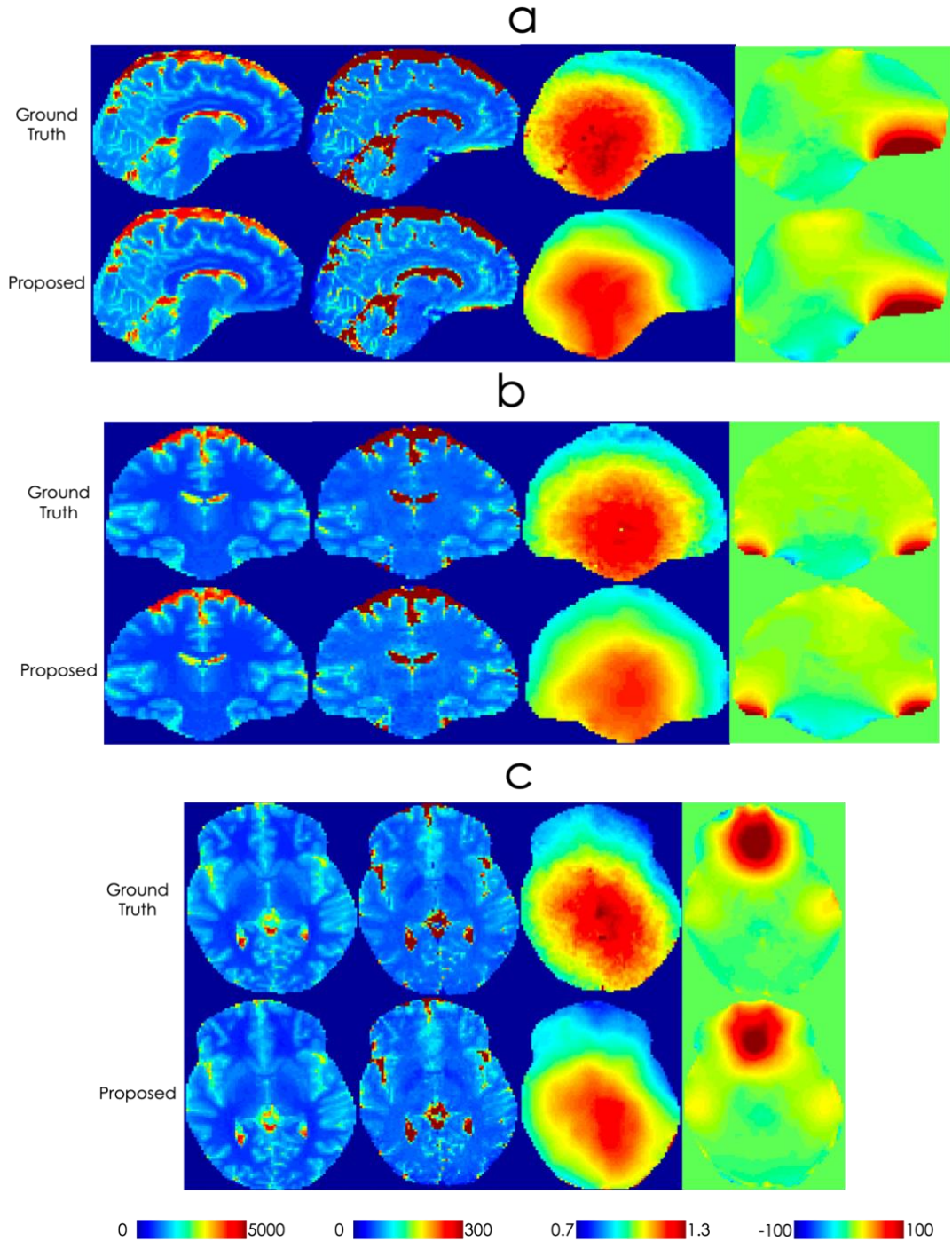

Figure 3. Orthogonal views of estimated parameter maps and field maps of sequence 1 using SAFE. Ground truth field maps are acquired by Physical calibration scan, and ground truth tissue parameter maps were acquired by $B_1^+$ & $B_0$-corrected MRF using ground truth field maps. For each view, from left to right, are T1, T2, $B_1^+$, $B_0$ maps respectively.

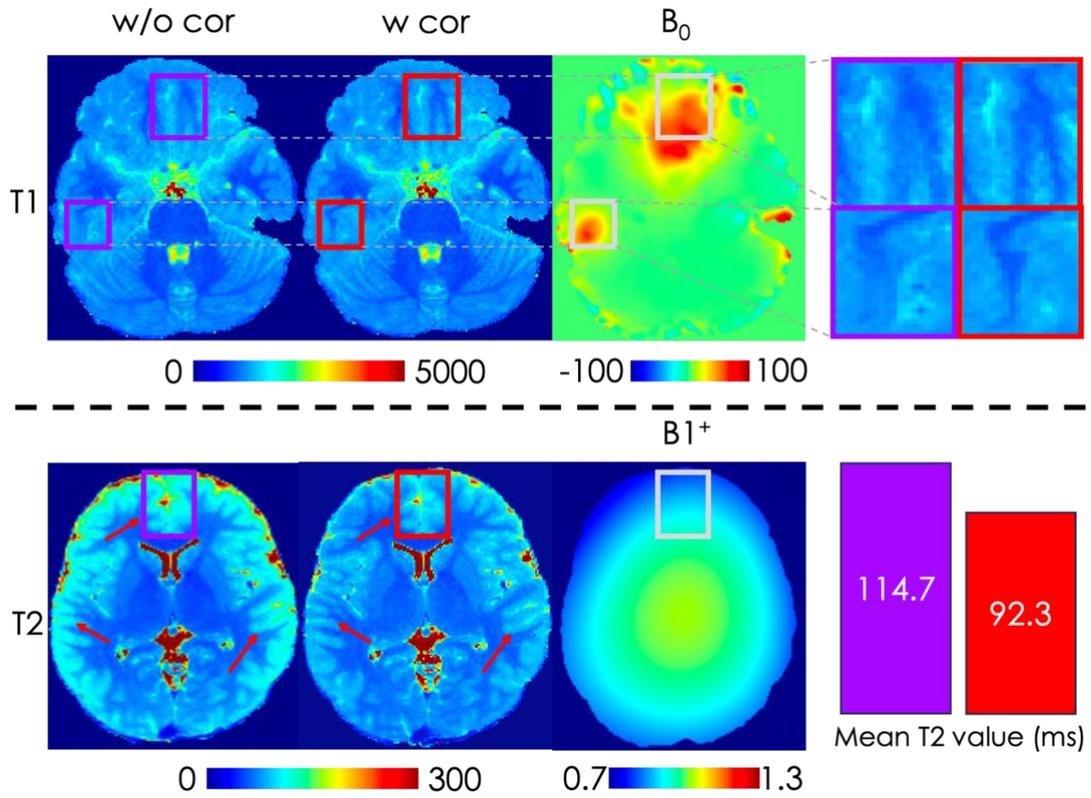

Figure 4. Application of SAFE to predict and correct for $B_1^+$ and $B_0$ inhomogeneities effects on previously acquired MRF data on children cohort. First row shows reduction in $B_0$-induced blurring on the T1 maps in the optical nerves and temporal lobe areas. Second row shows the T2 accuracy improvement around cingulate sulcus as a result of $B_1^+$-corrected dictionary-matching.

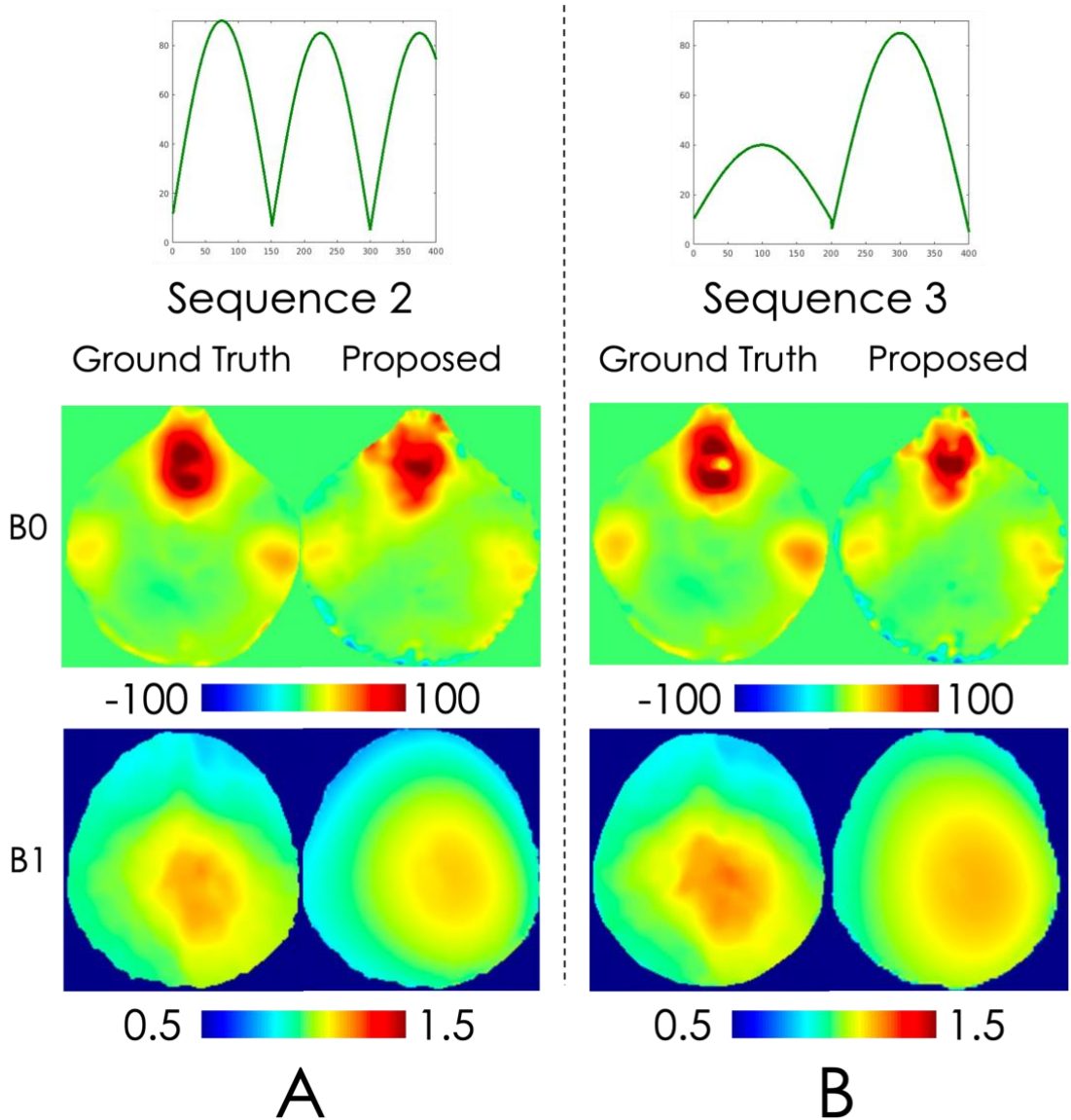

Figure 5. Two different MRF sequences were acquired to validate $B_1^+$ and $B_0$ estimation using SAFE, compared to the ground truth.